\documentclass[]{article}
\usepackage{graphicx,color}
\usepackage{amsmath,amssymb}
\usepackage[T1]{fontenc}
\usepackage{authblk}
\usepackage[framed]{mcode}
\usepackage{hyperref}

\newcommand{\D}{\mathrm{d}}

\newcommand{\Exp}[1]{\mathrm{e}^{\mbox{\footnotesize$#1$}}}
\newcommand{\I}{\mathrm{i}}
\newcommand{\tr}[1]{\mathrm{tr}{\left\{#1\right\}}}

\newcommand{\half}{\frac{1}{2}}

\begin{document}
\title{\textbf{Random samples of quantum states: Online resources\thanks{Please send your comments or suggestions to: QSampling@quantumlah.org.}}}

\author{Jiangwei {\sc Shang}$^{1,2}$,
Yi-Lin {\sc Seah}$^{1}$, Boyu {\sc Wang}$^{3}$, Hui Khoon {\sc Ng}$^{1,4,5}$,
David John {\sc Nott}$^{6}$ and Berthold-Georg {\sc Englert}$^{1,3,5}$
}
\affil{\small \emph{$^1$Centre for Quantum Technologies, National University of Singapore, %
Singapore 117543, Singapore} \\
\emph{$^2$Naturwissenschaftlich-Technische Fakult{\"a}t, Universit{\"a}t Siegen, 57068 Siegen, Germany}\\
\emph{$^3$Department of Physics, National University of Singapore, %
Singapore 117542, Singapore} \\
\emph{$^4$Yale-NUS College, Singapore 138614, Singapore} \\
\emph{$^5$MajuLab, CNRS-UNS-NUS-NTU International Joint Research Unit, UMI 3654, Singapore} \\
\emph{$^6$Department of Statistics and Applied Probability, National University of Singapore, %
Singapore 117546, Singapore}
}
\date{\vspace{-3ex} \small{(December 19, 2016)} \vspace{-2em}}
\maketitle
\begin{abstract}
This is the documentation for generating random samples from the quantum state space in accordance with a specified distribution, associated with this webpage\footnote{The full web address is \url{http://www.quantumlah.org/publications/software/QSampling/}.}: \url{http://tinyurl.com/QSampling}. Ready-made samples (each with at least a million points) from various distributions \cite{Shang+4:13,Shang+1:15,Seah+1:15} are available for download, or one can generate one's own samples from a chosen distribution using the provided source codes. The sampling relies on the Hamiltonian Monte Carlo algorithm as described in Ref.~\cite{Seah+1:15}. The random samples are reposited in the hope that they would be useful for a variety of tasks in quantum information and quantum computation. Constructing credible regions for tomographic data, optimizing a function over the quantum state space with a complicated landscape, testing the typicality of entanglement among states from a multipartite quantum system, or computing the average of some quantity of interest over a subset of quantum states are but some exemplary applications among many.
\end{abstract}

\newpage
\section{Introduction}
A measure of volume for regions in the quantum state space is needed in a range of situations, from comparing the relative importance of different regions, to quantifying the typicality of certain quantum properties. Any volume measure can be thought of as a probability distribution on the state space, with the volume of the full space normalized to unity. The choice of an appropriate measure or distribution depends on the operational situation at hand, and a variety have been used in the literature, e.g., the Hilbert--Schmidt measure, the Bures measure, the Haar measure on the eigenbasis choice complemented by a chosen distribution on the eigenvalues, etc. (see, for example, Chapter 14 in Ref.~\cite{GeomQStates}). 

Many of these distributions are easy to state conceptually, but are difficult to sample from, i.e., to produce a set of quantum states distributed in accordance with the chosen measure. Such samples are needed for any numerical calculation over the measure, e.g., computing the average value of a quantity over a given region. Here, we make use of Hamiltonian Monte Carlo (HMC) methods, adapted to the quantum situation as described in Ref.~\cite{Seah+1:15}.

As our work in Ref.~\cite{Seah+1:15} was motivated by quantum tomographic problems using a Bayesian approach, we provide ready-made samples for natural prior distributions for common tomography scenarios. The same HMC code can be adapted by users to generate samples relevant for their own purposes beyond tomography.

\section{Basic definitions}
\subsection{States, measurements, and probabilities}
The \emph{state} $\rho$ of a quantum system, also known as the statistical operator or the density matrix, is a nonnegative operator with unit trace, \textsl{i.e.},
\begin{equation}
  \rho\ge0 \quad\mbox{with}\,\,\tr{\rho}=1\,.
\end{equation}
In a finite $d$-dimensional Hilbert space, $\rho$ can be written as a $d\times d$ matrix, fully determined by $d^2-1$ independent real parameters.

A \emph{measurement} is described by a set of outcomes $\Pi\equiv \{\Pi_k\}_{k=1}^K$, where the $\Pi_k$s are nonnegative operators, $\Pi_k\geq 0$, with unit sum, $\sum_{k=1}^K\Pi_k=1$. $\Pi$ is referred to as a positive operator-valued measure (POVM), or a probability-operator measurement (POM). Each $\Pi_k$ corresponds to a detector in the measurement apparatus.

The probability that the $k$th detector clicks, given an input state $\rho$ to the measurement apparatus, is
\begin{equation}\label{eq:Born}
  p_k=\tr{\Pi_k\rho}\,,
\end{equation}
according to the Born rule. The nonnegative and unit-sum properties of the $\Pi_k$s assure that each $p_k\ge0$ and $\sum_k p_k=1$. We write $p\equiv \{p_k\}_{k=1}^K$ to denote the set of probabilities corresponding to the set of outcomes $\Pi$.

The POVM $\Pi$ can be thought of as a map from the state space of $\rho$s to the probability space of $p$s, $\Pi:\rho\longmapsto p$. $\Pi$ is generally a many-to-one map. A POVM $\Pi$ that is one-to-one, i.e., the probabilities $p$ identify a unique quantum state $\rho$, is said to be \emph{informationally complete} (IC); it is not informationally complete (NIC) otherwise.

In quantum tomography, it is more natural to work in the probability space, rather than in the state space: The tomographic data gathered for a chosen measurement $\Pi$ tell us about the probabilities $p$, not the state $\rho$; they are equivalent only for IC POVMs. Hence, the discussion below will be in the probability space. One can easily generalize to a different context where the state space is the natural arena: Choose a convenient IC POVM to establish a one-to-one mapping from the state space of interest to the probability space.

\subsection{Distributions and constraints}
We write the volume element of the infinitesimal vicinity of $p$ in the probability space as
\begin{equation}
(\D p)\, w(p)\,,
\end{equation}
with $w(p)$ being the target distribution of our choice, and $(\D p)$ denotes\begin{equation}
  (\D p)=\D p_1\D p_2\ldots \D p_K \, w_{\mathrm{cstr}}(p)\,.
\end{equation}
$w_{\mathrm{cstr}}(p)$ expresses the constraints imposed on the probability space by quantum mechanics: The nonnegative and unit-trace properties of $\rho$ translate---through $\Pi$---into constraints on $p$. $w_{\mathrm{cstr}}(p)$ is hence the indicator function for the subregion of the probability space relevant for quantum mechanics for a given $\Pi$. A $p$ is called \emph{physical} or \emph{permissible} if it satisfies all the constraints. For low-dimensional systems, $w_{\mathrm{cstr}}(p)$ can be written down explicitly; this becomes more difficult for high-dimensional systems, and is anyway not necessary for computational purposes (see Appendix A of Ref.~\cite{Shang+1:15}).

\section{Ready-made samples}
Here, we explain the various target distributions and POVMs used to create the samples found on the website. A list of samples available for download can be found in Table \ref{tbl:fileName}.

\subsection{Choice of target distribution}
The ready-made samples are generated in accordance with the following choices for the target distribution $w(p)$:
\begin{subequations}
\begin{align}
 \mbox{\bf Primitive prior}\!:&\,\, w_{\mathrm{prim}}(p)= 1\,,\\
 \mbox{\bf Jeffreys prior}\!:&\,\, w_{\mathrm{Jeff}}^{\ }(p)
 \propto\frac1{\sqrt{p_{1}p_{2}\cdots p_{K}}}\,,\\
 \mbox{\bf Conjugate prior}\!:&\,\, w_{\mathrm{conj}}(p)
 \propto p_1^{\beta_1}p_2^{\beta_2}\cdots p_K^{\beta_K}\,.
\end{align}
\end{subequations}
Here, the $w(p)$s are unnormalized---the normalization is inconsequential for the HMC algorithm as well as the samples generated. The terminology of a \emph{prior} is natural for our original intended Bayesian approach where a distribution $w(p)$ is chosen to encapsulate our knowledge about the probability space prior to data-taking; a user interested in other contexts should simply read ``prior'' as ``target distribution''.

The three specific priors are oft-encountered choices in quantum tomography. The primitive prior suggests that the distribution is uniform in $p$ over the
physical probability space.
The Jeffreys prior \cite{Jeffreys:46} is a common choice of prior when no external
prior information is available. The conjugate prior is a convenient way of incorporating prior knowledge that a particular $p$ (e.g., a particular $p$ is the target preparation) or region of the probability space is more likely than others. Here, we employ specifically the conjugate prior with
the hyperparameters set to $\beta_1=\beta_2=\cdots=\beta_K=1$.

\subsection{Various POVMs}
The POVMs used to generate the samples are common ones employed in tomography experiments. For contexts beyond tomography, the IC POVMs can be the chosen mapping between $\rho$ and $p$. Below, $\eta(~)$ refers to the Heaviside unit step function and $\delta(~)$ is the Dirac delta function.

\subsubsection{Single qubit ($d=2$) with IC POVM}
We consider two POVMs: the four-outcome tetrahedron POVM of minimal qubit tomography \cite{Rehacek+1:04}, and the six-outcome Pauli POVM that measures the three mutually unbiased bases of the Pauli matrices $\sigma_x$, $\sigma_y$ and $\sigma_z$. The tetrahedron POVM is a symmetric informationlly complete (SIC) POVM for dimension 2. In the following, we have ${x\equiv\langle\sigma_x\rangle}$, ${y\equiv\langle\sigma_y\rangle}$, and ${z\equiv\langle\sigma_z\rangle}$ for the expectation values of the Pauli matrices.
\begin{description}
 \item[Tetrahedron POVM] ($K=4$)
 \begin{eqnarray}\label{eq:tetra}
  p_{1} &=& \frac{1}{4}+\frac{1}{4\sqrt{3}}(x-y-z)\,,\quad
  p_{2}  =  \frac{1}{4}+\frac{1}{4\sqrt{3}}(y-z-x)\,,\nonumber \\
  p_{3} &=& \frac{1}{4}+\frac{1}{4\sqrt{3}}(z-x-y)\,,\quad
  p_{4}  =  \frac{1}{4}+\frac{1}{4\sqrt{3}}(x+y+z)\,.
\end{eqnarray}
The constraint factor $w_{\mathrm{cstr}}(p)$ is
\begin{equation}
  w_{\mathrm{cstr}}(p)=\eta(p_1)\,\eta(p_2)\,\eta(p_3)\,\eta(p_4)\,
\delta{\left(\sum\limits _{k=1}^4 p_k-1\right)}\,\eta{\left(\frac{1}{3}-\sum_{l=1}^4p_l^2\right)}.
\end{equation}
 \item[Pauli POVM] ($K=6$)
 \begin{eqnarray}\label{eq:Pauli}
  \left.\begin{array}{l}
  p_{1}\\
  p_{4}
 \end{array}\right\} = \frac{1}{6}(1\pm x)\,,\quad
  \left.\begin{array}{l}
  p_{2}\\
  p_{5}
 \end{array}\right\} = \frac{1}{6}(1\pm y)\,,\quad
 \left.\begin{array}{l}
  p_{3}\\
  p_{6}
 \end{array}\right\} = \frac{1}{6}(1\pm z)\,.
\end{eqnarray}
In this case, the constraint factor is
\begin{align}
 w_{\mathrm{cstr}}(p)&={\left[\prod_{k=1}^{3}\delta\Bigl(p_{k}+p_{k+3}-\frac{1}{3}\Bigr)\,\eta(p_{k})\,\eta(p_{k+3})\right]}\,\eta\biggl(\frac{1}{9}-\sum_{l=1}^{3}(p_{l}-p_{l+3})^{2}\biggr).
\end{align}
\end{description}

\subsubsection{Single qubit ($d=2$) with NIC POVM}
Here, we consider two NIC POVMs that provide no information about $y$.
\begin{description}
 \item[Trine POVM] ($K=3$)
  \begin{eqnarray}\label{eq:trine}
   p_1=\frac{1}{3}(1+z),\quad
   \begin{array}{l}p_2\\p_3\end{array}\biggr\}
   =\frac{1}{3}{\left(1-\frac{1}{2}z\pm\frac{\sqrt{3}}{2}\,x\right)}.
  \end{eqnarray}
   The constraint is
\begin{eqnarray}
  w_{\mathrm{cstr}}(p) &=& \eta(p_1)\,\eta(p_2)\,\eta(p_3)\,\delta{\left(\sum_{k=1}^3p_k-1\right)}\,\eta{\left(\half-\sum_{l=1}^3p_l^2\right)}.
\end{eqnarray}
 \item[Crosshair POVM] ($K=4$)
 \begin{eqnarray}\label{eq:cross}
  \left.\begin{array}{l}
  p_{1}\\
  p_{3}
 \end{array}\right\} = \frac{1}{4}(1\pm z)\,,\quad
 \left.\begin{array}{l}
  p_{2}\\
  p_{4}
 \end{array}\right\} = \frac{1}{4}(1\pm x)\,.
\end{eqnarray}
The constraint is
\begin{eqnarray}
 w_{\mathrm{cstr}}(p)&=&{\left[\prod_{k=1,2}\delta{\left(p_{k}+p_{k+2}-\half\right)}\,\eta(p_{k})\,\eta(p_{k+2})\right]}\nonumber\\
&&\qquad\times ~\eta{\left(\frac{1}{4}-(p_{1}-p_{3})^{2}-(p_{2}-p_{4})^{2}\right)}.
\end{eqnarray}
\end{description}

\subsubsection{Qutrit ($d=3$) with IC POVM}
In dimension 3, there exists a one-parameter family of nonequivalent SIC POVMs. We choose one of them, written in vector form (the columns), such that,
\begin{equation}\label{eq:qutritSIC}
\left[
  \begin{array}{c@{\hspace{.8em}}c@{\hspace{.8em}}c@{\hspace{.8em}}c@{\hspace{.8em}}
  c@{\hspace{.8em}}c@{\hspace{.8em}}c@{\hspace{.8em}}c@{\hspace{.8em}}c}
    1      & 1             & 1 & 0      & 0             & 0 & \omega  & \omega^{\ast} & 1 \\
    \omega & \omega^{\ast} & 1 & 1      & 1             & 1 & 0       & 0             & 0 \\
    0      & 0             & 0 & \omega & \omega^{\ast} & 1 & 1       & 1             & 1 \\
  \end{array}
\right]\!,
\end{equation}
where $\omega=\Exp{\I 2\pi/3}$ is the cubic root of unity, with $\omega^*=\omega^2$ and $1+\omega +\omega^2=0$.

\subsubsection{Two qubits ($d=4$) with IC POVM}
We consider the scenario where each of the two qubits is measured by the four-outcome tetrahedron POVM of Eq.~\eqref{eq:tetra} respectively. The resulting two-qubit POVM (which is IC) has sixteen outcomes with the single $\delta$-function constraint of unit sum, so the probability space is 15-dimensional.

\subsubsection{Two qubits ($d=4$) with NIC POVM}
In the BB84 scenario of quantum key distribution, the two parties use two four-outcome crosshair measurements of Eq.~\eqref{eq:cross} and so have sixteen outcomes for the composed POVM. However, there are eight $\delta$-function constraints for the probabilities, resulting in the probability space being 8-dimensional.

On the other hand, one qubit can be measured by using the three-outcome trine measurement of Eq.~\eqref{eq:trine}, and another by the anti-trine measurement with the signs of $z$ and $x$ changed in Eq.~\eqref{eq:trine}, namely, the TAT scheme. The resulting POVM has nine outcomes with the single $\delta$-function constraint of unit sum, so the probability space is again 8-dimensional.

\subsubsection{Three qubits ($d=8$) and four qubits ($d=16$) with IC POVMs}
For higher dimensions, such as $d=8$ and $d=16$, the samples are generated using the following procedure (see footnote 11 in Ref.~\cite{Shang+1:15}): First generate a square random matrix $A$ with all entries being independent complex Gaussian numbers, then the quantum state is constructed by $\rho=AA^{\dag}/\tr{AA^{\dag}}$. This gives the primitive prior for any IC POVM. The method can be used in place of the HMC code (see next section) to generate samples according to the primitive prior for any dimension.

\subsection{Use of the samples}\label{sec:use}
The available samples are listed in Table~\ref{tbl:fileName}. There are two different formats for the sample files, namely, \textsl{.mat} and \textsl{.txt}. The \textsl{.mat} files are workspace files that can be directly loaded into \textsl{MATLAB}, with the variables explained in Table~\ref{tbl:check}. The \textsl{.txt} files store the real and imaginary parts (as two separate files) of the generated quantum states. Each row in the \textsl{.txt} files corresponds to one state, which can be easily formatted to a $d\times d$ matrix row after row. 

Different sets of POVMs define the same probability space if they are linearly related, so that the Jacobian for the transformation between the corresponding probabilities is a numerical value. Therefore, if the target distribution is the primitive prior (i.e., uniform over the physical probability space), it does not matter which POVM we choose for the sampling. We denote such cases by ``$\ast$'' in Table~\ref{tbl:fileName}.

The two-qubit quantum states generated with NIC POVMs carry different weights (corresponding files with suffix ``\textsl{\_range}''), because of the NIC nature of the POVMs used. For more details, please see Ref.~\cite{Seah+1:15}. In addition, we also provide the files of random samples with weights already integrated into the distribution by re-sampling the samples with respect to their weights. These files carry the suffix ``\textsl{\_w}''.

\begin{table}[t!]
\begin{changemargin}{-1cm}{-1cm}
\caption{\label{tbl:fileName}%
Available random samples. IC: informationally complete; NIC: not informationally complete; SIC: symmetric informationally complete. Since the primitive prior distribution is uniform in $p$ over the physical probability space, it does not matter which POVM we choose for the sampling, as long as all the POVMs (denoted by ``$\ast$'') define the same probability space. For more details, refer to Sec.~\ref{sec:use} and Ref.~\cite{Seah+1:15}.}
\vspace{-.2cm}
\begin{center}
\begin{tabular}{@{\hspace{.6em}}c|@{\hspace{.8em}}c|@{\hspace{.8em}}c|
@{\hspace{.8em}}c|@{\hspace{.8em}}c@{\hspace{.6em}}}
\hline\hline\rule{0pt}{14pt}
{\bf Files} & {\bf Dimension} & {\bf Distribution} & {\bf POVM} & {\bf IC or NIC} \\
\hline\rule{0pt}{14pt}
\verb|1qb_IC_prim| & $d=2$ & Primitive & $\ast$ & IC \\[1ex]
\verb|1qb_Jeff_tthd| & $d=2$ & Jeffreys & Tetrahedron, {Eq.~\eqref{eq:tetra}} & IC \\[1ex]
\verb|1qb_Jeff_Pauli| & $d=2$ & Jeffreys & Pauli, {Eq.~\eqref{eq:Pauli}} & IC \\[1ex]
\verb|1qb_conj_tthd| & $d=2$ & Conjugate & Tetrahedron, {Eq.~\eqref{eq:tetra}} & IC \\[1ex]
\verb|1qb_conj_Pauli| & $d=2$ & Conjugate & Pauli, {Eq.~\eqref{eq:Pauli}} & IC \\[1ex]
\hline\rule{0pt}{14pt}
\verb|1qb_NIC_prim| & $d=2$ & Primitive & $\ast$ & NIC \\[1ex]
\verb|1qb_Jeff_trine| & $d=2$ & Jeffreys & Trine, {Eq.~\eqref{eq:trine}} & NIC \\[1ex]
\verb|1qb_Jeff_cross| & $d=2$ & Jeffreys & Cross, {Eq.~\eqref{eq:cross}} & NIC \\[1ex]
\verb|1qb_conj_trine| & $d=2$ & Conjugate & Trine, {Eq.~\eqref{eq:trine}} & NIC \\[1ex]
\verb|1qb_conj_cross| & $d=2$ & Conjugate & Cross, {Eq.~\eqref{eq:cross}} & NIC \\[1ex]
\hline\rule{0pt}{14pt}
\verb|qutrit_prim_SIC| & $d=3$ & Primitive & SIC, {Eq.~\eqref{eq:qutritSIC}} & IC \\[1ex]
\verb|qutrit_Jeff_SIC| & $d=3$ & Jeffreys & SIC, {Eq.~\eqref{eq:qutritSIC}} & IC \\[1ex]
\verb|qutrit_conj_SIC| & $d=3$ & Conjugate & SIC, {Eq.~\eqref{eq:qutritSIC}} & IC \\[1ex]
\hline\rule{0pt}{14pt}
\verb|2qb_IC_prim| & $d=4$ & Primitive & $\ast$ & IC \\[1ex]
\verb|2qb_Jeff_2tthd| & $d=4$ & Jeffreys & Two tetrahedrons & IC \\[1ex]
\verb|2qb_conj_2tthd| & $d=4$ & Conjugate & Two tetrahedrons & IC \\[1ex]
\hline\rule{0pt}{14pt}
\verb|2qb_NIC_prim| & $d=4$ & Primitive & $\ast$ & NIC \\[1ex]
\verb|2qb_Jeff_TAT| & $d=4$ & Jeffreys & TAT & NIC \\[1ex]
\verb|2qb_Jeff_BB84| & $d=4$ & Jeffreys & BB84 & NIC \\[1ex]
\verb|2qb_conj_TAT| & $d=4$ & Conjugate & TAT & NIC \\[1ex]
\verb|2qb_conj_BB84| & $d=4$ & Conjugate & BB84 & NIC \\[1ex]
\hline\rule{0pt}{14pt}
\verb|3qb_IC_prim| & $d=8$ & Primitive & $\ast$ & IC \\[1ex]
\hline\rule{0pt}{14pt}
\verb|4qb_IC_prim| & $d=16$ & Primitive & $\ast$ & IC \\[1ex]
\hline\hline
\end{tabular}
\end{center}
\end{changemargin}
\end{table}
\begin{table}[t!]
\caption{\label{tbl:check}%
Interpretation of the symbols contained in the \textsl{.mat} files.
The \textsl{.txt} files only contain the random quantum states generated
(also weights for the two-qubit states with NIC POVMs).}
\vspace{-.2cm}
\begin{center}
\begin{tabular}{@{\hspace{.6em}}c|@{\hspace{.8em}}c@{\hspace{.6em}}}
\hline\hline\rule{0pt}{14pt}
{\bf Symbols} & {\bf Interpretation} \\
\hline\rule{0pt}{14pt}
\verb|rho| & quantum states generated \\[1ex]
\verb|purity| & purity of quantum states \\[1ex]
\verb|range| & weight for two-qubit states with NIC POVMs \\[1ex]
\verb|acceptrate| & HMC acceptance rate \\[1ex]
\hline\rule{0pt}{14pt}
\verb|d| & dimension of Hilbert space \\[1ex]
\verb|sigma| & Pauli operators \\[1ex]
\verb|pom| & POVM \\[1ex]
\verb|numstep| & number of states generated  \\[1ex]
\hline\rule{0pt}{14pt}
\verb|nt| & number of $\theta$ parameters in HMC \\[1ex]
\verb|nf| & number of $\phi$ parameters in HMC \\[1ex]
\verb|num| & $\equiv$ nt+nf, total number of independent parameters \\[1ex]
\verb|pvar| & step size in momentum space \\[1ex]
\verb|qvar| & step size in position space \\[1ex]
\verb|nint| & number of jumps in each leapfrog run \\[1ex]
\verb|stepsize| & $\equiv$ qvar/nint, step size in each jump \\[1ex]
\hline\hline
\end{tabular}
\end{center}
\end{table}

\clearpage

\section{Generate your own samples}
The samples listed in Table~\ref{tbl:fileName} are generated by the corresponding source codes, namely the ``\textsl{hmc\_}'' files. To generate your own samples in accordance with any distribution of choice, refer to the source codes ``\textsl{hmc.m}'' and ``\textsl{spect.m}''. 

As an example of how to use the code, consider $\verb|2qb_conj_2tthd|$ of Table~\ref{tbl:fileName}. The corresponding \textsl{MATLAB} scripts are copied below. One needs to be specify the dimension $d$, the POVM outcomes, and the target distribution.

Note that the current HMC code may yield inaccurate results beyond $d=5$. While the HMC method works, in principle, for any dimension, numerical issues with the current \textsl{MATLAB} implementation---due to the extreme small size of the Jacobian used in the code when $d$ gets large---have been observed to occur for dimensions $d=6$ and up when the code is run on a standard desktop machine.

\begin{changemargin}{-1cm}{-1cm}
\begin{lstlisting}
%--------------------------------- hmc.m ---------------------------------%
%
%--------------- hmc settings ---------------%
numstep=1000000; % number of states generated
pvar=1; % step size in momentum space
qvar=0.1; % step size in position space
nint=10; % number of jumps in each leapfrog run
stepsize=qvar/nint; % step size in each jump

%--------------- spect decomp settings ---------------%
d=4; % dimension of Hilbert space
num=d^2-1; % total number of indpendent parameters

%-------------------------------------------------------------------------%
%--------------- POVM settings ---------------%
sigma=zeros(2, 2, 4); % Pauli matrices
sigma(:,:,1)=[1 0; 0 1]; % identity
sigma(:,:,2)=[1 0; 0 -1]; % sigma_z
sigma(:,:,3)=[0 1; 1 0]; % sigma_x
sigma(:,:,4)=[0 -1i; 1i 0]; % sigma_y

T=zeros(2,2,4); % Tetrahedron POVM
T(:,:,1)=sigma(:,:,1)+(sigma(:,:,2)-sigma(:,:,3)-sigma(:,:,4))/sqrt(3);
T(:,:,2)=sigma(:,:,1)+(-sigma(:,:,2)+sigma(:,:,3)-sigma(:,:,4))/sqrt(3);
T(:,:,3)=sigma(:,:,1)+(-sigma(:,:,2)-sigma(:,:,3)+sigma(:,:,4))/sqrt(3);
T(:,:,4)=sigma(:,:,1)+(sigma(:,:,2)+sigma(:,:,3)+sigma(:,:,4))/sqrt(3);

Q=zeros(d,d,d^2); % overall POVM: Two tetrahedrons
for j=1:d^2
    Q(:,:,j)=kron(T(:,:,ceil(j/d)),T(:,:,1+rem(j-1,d)));
end

pom=Q(:,:,1:num); % reduced POVM: independent POVM elements only,
% corresponding to the same number of independent parameters used in HMC

\end{lstlisting}
\newpage
\begin{lstlisting}
%-------------------------------- spect.m --------------------------------%
%
%--------------- distribution ---------------%
prior=abs(prod(p_full)); % conjugate prior, for instance
% p_full corresponds to Q, the overall POVM

u_prior=zeros(1,num); % gradient invoked from prior distribution
for i=1:num
    for j=1:d^2
        u_likeli(i)=u_likeli(i)+D(j)/p_full(j)*dpdm_full(j,i); % for conjugate prior
        % dpdm_full is the first derivative of p_full w.r.t. the angle parameters
    end
end

\end{lstlisting}
\end{changemargin}

\section*{Acknowledgement}
This work is funded by the Singapore Ministry of Education (partly through
the Academic Research Fund Tier 3 MOE2012-T3-1-009) and the National Research
Foundation of Singapore. H. K. Ng is also funded by a Yale-NUS College start-up grant.

%

\end{document}